\documentclass[twocolumn,pre,showpacs,superscriptaddress]{revtex4}
\usepackage{color}
\usepackage{graphicx}%
\usepackage{psfrag}
\usepackage{dcolumn}
\usepackage{amsmath}
\usepackage{latexsym}

\begin{document}

\def\clr{\color{red}}
\def\bea{\begin{eqnarray}}
\def\eea{\end{eqnarray}}
\def\beq{\begin{equation}}
\def\eeq{\end{equation}}
\def\f{\frac}
\def\k{\kappa}
\def\sx{\sigma_{xx}}
\def\sy{\sigma_{yy}}
\def\sxy{\sigma_{xy}}
\def\e{\epsilon}
\def\ve{\varepsilon}
\def\ex{\epsilon_{xx}}
\def\ey{\epsilon_{yy}}
\def\exy{\epsilon_{xy}}
\def\b{\beta}
\def\D{\Delta}
\def\h{\theta}
\def\t{\tau}
\def\r{\rho}
\def\trz{\rho^{(0)}}
\def\trek{\rho^{(1)}}
\def\tcz{C^{(0)}}
\def\a{\alpha}
\def\s{\sigma}
\def\kb{k_B}
\def\la{\langle}
\def\ra{\rangle}
\def\nn{\nonumber}
\def\bu{{\bf u}}
\def\bn{\bar{n}}
\def\br{{\bf r}}
\def\up{\uparrow}
\def\dn{\downarrow}
\def\S{\Sigma}
\def\dg{\dagger}
\def\d{\delta}
\def\p{\partial}
\def\l{\lambda}
\def\G{\Gamma}
\def\o{\omega}
\def\g{\gamma}
\def\kv{\bar{k}}
\def\ha{\hat{A}}
\def\hv{\hat{V}}
\def\hg{\hat{g}}
\def\hG{\hat{G}}
\def\hTT{\hat{T}}
\def\noi{\noindent}
\def\a{\alpha}
\def\d{\delta}
\def\p{\partial} 
\def\r{\rho}
\def\xv{\vec{x}}
\def\rv{\vec{r}}
\def\fv{\vec{f}}
\def\ov{\vec{0}}
\def\vv{\vec{v}}
\def\la{\langle}
\def\ra{\rangle}
\def\e{\epsilon}
\def\o{\omega}
\def\n{\eta}
\def\g{\gamma}
\def\th{\hat{t}}
\def\uh{\hat{u}}
\def\break#1{\pagebreak \vspace*{#1}}
\def\f{\frac}
\def\hf{\frac{1}{2}}
\def\uu{\vec{u}}
\newcommand{\be}{\begin{equation}}
\newcommand{\ee}{\end{equation}}
\newcommand{\eps}{\epsilon}
\newcommand{\w}{\omega}

\bibliographystyle{apsrev4-1}
\title{Bimodal response in periodically driven diffusive systems}



\author{Urna Basu}
\email{urna.basu@saha.ac.in} 
\affiliation{
TCMP Division, Saha Institute of Nuclear Physics, 1/AF Bidhan Nagar,
Kolkata 700064, India
}
\author{Debasish Chaudhuri}
\email{d.chaudhuri@amolf.nl}
\affiliation{FOM Institute AMOLF,  Science Park 104, 1098 XG Amsterdam, The Netherlands}
\author{P. K. Mohanty}
\affiliation{
TCMP Division, Saha Institute of Nuclear Physics, 1/AF Bidhan Nagar,
Kolkata 700064, India
}


\date{\today}

\begin{abstract}
{
We study the response of one dimensional diffusive systems,  
 consisting of particles interacting via 
symmetric or asymmetric exclusion, to time-periodic 
driving from two reservoirs coupled to the ends. The dynamical response of the system can be
characterized in terms of  the structure factor.
We find an interesting frequency dependent response  -- 
the  current carrying majority excitons cyclically crosses over
from a short wavelength mode to a long wavelength mode with an intermediate
regime of coexistence. This effect being boundary driven, decays inversely with system size.
Analytic calculations show that this
behavior is common to diffusive systems,  both in absence and presence of correlations.
}
\end{abstract}

\maketitle

\section{Introduction}
Study of one dimensional driven diffusive systems (DDS) has drawn a  lot of attention in recent years\cite{book}. 
Exclusion processes \cite{Exclusion} have been studied extensively as  the paradigm models 
of DDS. Despite being simple,   they exhibit rich variety of phases, 
exotic correlations,  non-trivial variations of density-profiles and  currents providing 
deep understanding  of  non-equilibrium transport.
These non-equilibrium systems  are usually driven  by coupling them  to reservoirs at the boundaries. 
The simplest example of DDS is a symmetric exclusion process (SEP) where  hardcore particles
on a one dimensional lattice hop to any of  their nearest neighbors, if vacant, with equal rates. 
Open system of SEP driven by time-independent bias at the boundaries has been studied in 
details~\cite{Derrida2002, Derrida2004, Santos2001}; its density profiles~\cite{Derrida2002}, 
correlation functions~\cite{Santos2001}, and current fluctuations~\cite{Derrida2004} have been
calculated. DDS with  asymmetric bulk dynamics, the asymmetric simple exclusion
process (ASEP) where particles  preferably hop to one direction and the  totally asymmetric 
simple exclusion process (TASEP) where particles can  only hop  forward 
have also been studied extensively with time independent bias at the boundaries\cite{tasep1}.

Systems with time dependent bias at the boundaries are also studied in different contexts. 
The most commonplace example of periodically driven diffusive system can be found in alternating current (AC) 
driven classical electrical circuits where electron transport is best described as Drude diffusion. 
Studies involving 
classical stochastic system of particles under the influence of  time-periodic potentials~\cite{Jung}
remained an active field that led to interesting phenomena like stochastic resonance~\cite{Gammaitoni1998}.
In recent years, this field has received renewed interest in the context of
Brownian ratchet models of motor-proteins~\cite{Julicher1997, Astumian2002}, ion-pumps associated with 
cell membrane~\cite{Astumian2003}, and classical particle-pump models~\cite{Jain2007, Marathe2008}.
AC driven TASEP is recently studied in the 
context of transport of vehicular traffics on modern highways under the influence of 
periodically modulating red/green light signal phases~\cite{Popkov2008}.

In this paper,  we focus on  characterizing a  system of one dimensional 
diffusing interacting particles   driven by
two reservoirs   coupled to the system at its two ends.
We
characterize the response of this system  in terms of structure factor.
We find an interesting bi-modality of the majority excitons. At a given frequency, the majority excitons
cyclically crosses over from a predominantly long wavelength mode  to a short wavelength mode via an
intermediate regime of bi-modality. Our analytic calculations show that this behavior, which stems from 
the diffusive nature of the system, persists even   in presence of finite inter-particle correlations.

\section{Model: periodic drive}
 
The asymmetric simple exclusion process (ASEP) is defined on a one dimensional 
lattice  where sites are labelled  
by $i=0,1,2\dots N$. Each site can either be vacant or occupied by at most one  
particle; the corresponding site variables are $n_i=0$  and $n_i=1$  respectively. 
The particles can hop to  the right (left) nearest neighbouring site, if vacant, 
with rate $p$ ($q$). A special case of the dynamics  with $p=q$ denotes SEP. The  local density 
$\r_i = \la n_i \ra$  and other correlation functions  of the system, thus evolve  according 
to the following particle conserving dynamics (in the bulk),   
\bea
1~ 0 \, \mathop{\rightleftharpoons}_{q}^{p} \, 0 ~1.
\eea

At the boundaries, the system is coupled to two  reservoirs, which effectively 
fixes the density of the boundary sites $i=0,N.$  The time varying drive,  in  
density is introduced as 
\bea
 i=0~ {\rm (left)}  &~:~~~&  \rho_0   =  \rho_L + a(t) \cr
 i=N ~{\rm (right)} &~:~~~& \rho_N  =  \rho_R - a(t) 
\eea
For simplicity we consider  sinusoidal variation $a(t)=\a\sin\w t$  all through this study.

The time-evolution of density profile is then given by 
\bea
\frac{d\r_i}{dt}  &=& J_{i-1,i} - J_{i,i+1} \cr
	  &=& p (\r_{i-1}- \r_i + C_{i,i+1} -C_{i-1,i}) \cr
&& + q (\r_{i+1} - \r_i +C_{i-1,i} - C_{i,i+1})
\label{eq:rhoi}
\eea 
where $J_{i,i+1} = p\, \r_i - q\, \r_{i+1} - (p-q) C_{i,i+1}$ is the  current
through $i$-th bond and  $C_{i_1,i_2\dots i_k} =\la n_{i_1} n_{i_2} \dots   n_{i_k}\ra$  
are the  $k$-point correlation functions. Note that  Eq. \eqref{eq:rhoi}   involves  
the $2$-point correlations which  themselves evolve as  
\bea
 \frac{d}{dt}{ C_{i,i+1}} &=& p C_{i-1,i+1} + q C_{i,i+2} - (p+q) C_{i,i+1}\cr
&& + (p-q) (C_{i,i+1,i+2} - C_{i-1,i,i+1}).  \nn
\eea
Again  the above equation involves $3$-point correlations. In general, time evolution of 
$k$-point correlation  depends on $(k+1)$-point correlations and thus  it is difficult to 
solve these hierarchical set of equations.  However, for SEP ($p=q$) time evolution of 
local density (Eq. \eqref{eq:rhoi}) gets decoupled from $2$-point correlations. 
Thus, we begin our discussion by analysing this simplest case of $p=q$ first, 
before going into the  details of the generic case $p\neq q$.

\section{SEP with AC drive}
SEP in the bulk ($p=q \equiv r$) leads to a simple diffusive time evolution of the 
density profile 
\be
\frac{d\rho_i}{dt}=r\, \left(\rho_{i+1} + \rho_{i-1} - 2\rho_i\right)~~~~~\mbox{for} ~i=1,\dots,N-1.  
\label{eq:ro}
\ee
For DC bias $a(t)=0$ and $\r_L \neq \r_R$, the steady 
state density profile is linear
$\rho_i^s = \rho_L - (\rho_L-\rho_R)\frac{i}{N}$.  
To focus our attention entirely on AC response we set $\r_L=\r_R=\r$. 
The requirement that the occupation at each site is bounded between zero and one, 
leads to the following bound to the amplitude of the oscillatory density at the boundaries 
$\a\leq {\rm min}[\r,\,(1-\r)]$.
The resulting density profile is
\beq 
\r_i=\r+\phi_i(t),
\label{eq:rho_i}
\eeq
 where $\phi_i(t)$
is a function arising entirely due to the AC driving.
To get rid of the time dependent boundary conditions let us define a 
new variable   
$     
u_i(t)=\phi_i(t) + f_i a(t),
$
such that $u_0=0=u_N$.
A linear function of $i$,  $f_i=(2i/N-1)$ maintains these criteria.  
Thus the time evolution of
$u_i$, 
\bea\label{eq:u}
\frac{du_i}{dt}=r\, (u_{i-1}+u_{i+1}-2 u_i)+f_i \frac{da(t)}{dt}, ~ 1\leq i \leq N-1.
\eea
This  set of coupled differential equations can be easily 
decoupled  by the discrete Fourier sine transform (as $u_0=0=u_N$)
\bea
u_i(t)=\sum_{j=0}^{N} \tilde{u}_j(t)~ \psi_j(i) 
\eea
where $\psi_j(i)= \sin (k_j i )$ with $j$ or $k_j=\pi j/N$ denoting the sine Fourier modes.
In this basis Eq.~\eqref{eq:u} becomes (using the fact $a(t)= \alpha \sin \omega t$)
\bea\label{eq:uk}
\frac{d}{dt}\tilde{u}_j(t)=\e_j \tilde u_j(t) + \tilde f_j \a \w \cos \w t
\eea
where $\e_j=2r\,(\cos k_j-1)$  and 
$\tilde f_j$ is the discrete sine transform of $f_i$, i.e.,
\bea
\tilde f_j &=& \frac{2}{N} \sum_{i=1}^{N-1}f_i~ \psi_j(i) \nn\\
	   &=& -\frac{[1+(-1)^j]}{N} \cot \left(\frac{k_j}{2}\right). 
\eea
Eq.~\eqref{eq:uk} has the general solution
\bea
\tilde u_j(\w,t) &=& \tilde u_j(0) e^{\e_j t}\nn\\ 
&+& \f{\a \w \tilde f_j}{\e_j^2+\w^2}\left(-\e_j \cos \w t+\w \sin \w t +\e_j e^{\e_j t}\right).
\eea
Clearly the relaxation time of $j$-th mode is $\t_j=-1/\e_j$, the longest
relaxation time being $\t\equiv \t_{{j=2}}=(1/r)(N/2\pi)^2$. $\t$ is interpreted as the
relaxation time of the system.
In the limit  of $t\gg \t$ the  dependence on initial condition $\tilde u_j(0)$
drops out (as $\e_j\leq 0$) 
and we are left with a temporally periodic function of  $\h=\w t$, 
%
\begin{figure}[t]
\includegraphics[width=7cm]{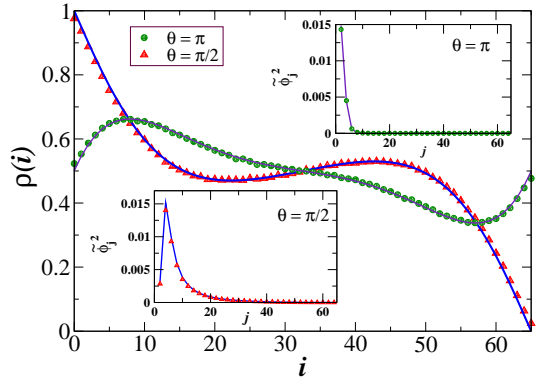}
\caption{ (Color online) Comparison of analytical density profiles \eqref{eq:rho_i} and corresponding structure factors 
\eqref{eq:phit_j} (shown in the insets) with those obtained from the Monte-Carlo simulations (symbols).
The parameters are  $N=65$, $r=1/2$, $\rho=0.5$, $\a=0.5$ and  $\o=0.01.$}
\label{fig:rho_i}
\end{figure}
\bea
\tilde u_j(\o,\h) = \frac{\a \o \tilde f_j}{\e_j^2+\o^2}\left(-\e_j \cos \h+\o \sin \h \right).
\eea
The  AC response in density profile in the long time  limit,  can be  obtained from the Fourier sum,
\bea
\phi_i(\omega,\h) = \sum_{j=1}^{N-1} \tilde{\phi}_j(\o,\h)~ \psi_j(i)
\label{eq:phi_i}
\eea
where the Fourier modes are
\bea
\label{eq:mode}
\tilde\phi_j(\o,\h) & =& \tilde u_j -\tilde f_j a(t) \nn\\ 
 &=& -\f{2\a}{N}\f{\sin k_j}{\e_j^2+\w^2}\left(\w \cos \h + \e_j \sin \h\right)	\label{eq:phit_j}
\eea
for $j=~$even integers, and $\tilde\phi_j(\o,\h) =0$ for $j=~$odd integers.
The profile $\phi_i(\o,\h) $ is independent of initial conditions and periodic in 
$\h=\w t$, a feature typical of periodically driven systems.  
The density profile $\rho_i$  for any given $\omega$ and  $\h$ can   be calculated now using 
Eq.  \eqref{eq:phi_i}  and \eqref{eq:mode} in  Eq. \eqref{eq:rho_i}.

\begin{figure}[t]
\psfrag{N2phi2}{$\left(N\tilde\phi_j/2\a \right)^2$}
\psfrag{k}{$k_j$}
\psfrag{q=.35    }{$\h=0.35$}
\psfrag{q=.88    }{$\h=0.88$}
\psfrag{q=1.2    }{$\h=1.2$}
\psfrag{q=1.6    }{$\h=1.6$}
 \includegraphics[width=8.5cm]{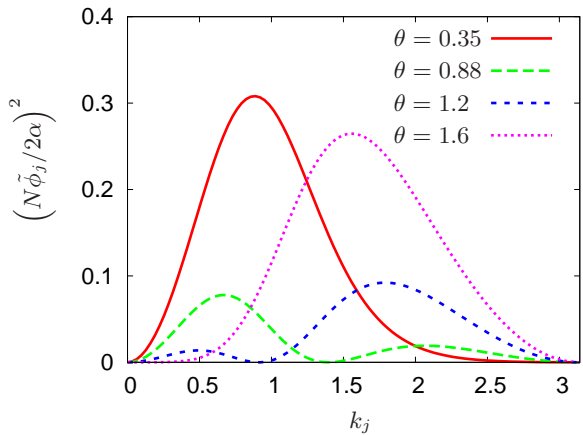}
\caption{(Color online)
The normalized structure factor $(N\tilde\phi_j/2\a)^2$ as a function of 
$k_j=\pi j/N$ evaluated at various phases $\h$
(indicated in the legend) and frequency $\w=1$. We used hopping rate 
$r=1/2$. 
}
\label{phik}
\end{figure}

In Fig. ~\ref{fig:rho_i}  we compare  this analytically obtained density profile $\rho_i$ (main figure)
and  $\tilde\phi_j^2$  (insets) for a system of size $N=65$ with those obtained from Monte-Carlo 
simulations for two different values of $\h=\pi/2,\pi$.  The amplitude  and frequency of the 
AC drive   are taken as $\a=0.5$ and  $\o=0.01$ respectively, and  
 the mean density is kept fixed at $\rho=0.5$.

The structure factor (square of the sine Fourier modes   corresponding to the AC response)   
$\tilde \phi_j^2$   is a function of $\w$ and $\h$,
such that $\tilde \phi_j^2(\omega,\h+\pi) =\tilde \phi_j^2(\omega,\h).$ 
So the relevant physical domain is 
$0<\w<\infty$, $0 \leq \h \leq \pi$. In the following we 
 analyze the   behaviour of 
the   dominating  mode(s)  of the structure factor in this domain. 

It is clear from Eq. \eqref{eq:mode} that  
for any $\w>0$, $\tilde\phi_j= 0$ at $k_j =0$ and  $\pi$, 
indicating   the presence of {\em at least one} maximum  in $\tilde\phi_j^2$ at 
some  non-zero $k_j<\pi.$
However, $\tilde \phi_j$ has another {\em zero} 
when $( \omega \cos \h +\epsilon_j \sin \h)$ vanishes 
resulting in an additional minimum in $\tilde\phi_j^2$. 
This minimum occurs at a  specific mode $j$  for
which $\e_j= - \frac{\w}{\tan \h}.$ 
Note that the system can access this mode   
only when  
\bea
\tan^{-1} \left(\frac{\omega}{4r}\right) \leq \h \leq \frac{\pi}{2}, 
\label{eq:bimod}
\eea
because  $\e_j$ is bounded within the range $-4r\leq\e_j\leq 0.$
Thus, in this regime of $\theta$ 
 the structure factor  will have {\em at least two} maxima.

\begin{figure}[t]
\psfrag{q}{$\h$}
\psfrag{w}{$\w$}
\psfrag{kg}{$k_>$}
\psfrag{kl}{$k_<$}
\psfrag{single-mode}{single-mode}
\psfrag{bimodal}{bimodal}
 \includegraphics[width=8.5cm]{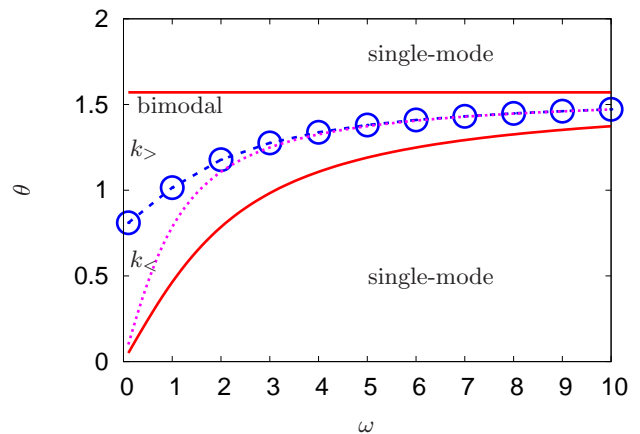}
\caption{(Color online) For SEP with hopping rates $r=1/2$
bi-modality is possible in the range $\tan^{-1}(\w/2)\leq \h\leq \pi/2$
denoted by the solid lines. The dashed line through circles denote the 
{\em line of crossover} at which the two strongest modes become of equal 
strength. Only one strong mode is operative in the region denoted by 
single-mode. By $k_>$  ($k_<$) we denote the regime where the large $k$
(small $k$) mode is stronger. In the limit
of large $\w$ the line of cross-over can be analytically found to be 
$\tan \h =\w$ and is shown as the dotted line.
}
\label{phd}
\end{figure}

Clearly  there is a distinct  region in the  $\o$-$\h$ plane, given by Eq. \eqref{eq:bimod}, where 
$\tilde\phi_j^2$ has  bimodal structure with two dominating  modes (Fig. \ref{phik}). 
Outside this region the structure factor will remain unimodal, as shown in Fig.~\ref{phd}. 
At any given frequency $\w$, the typical nature of  the structure  factor
undergoes the following transformations: with increasing $\h$ the structure factor changes
from unimodal to a bimodal  behaviour with the second peak
appearing at a higher value of $k_j$ or shorter wavelength. Eventually at a larger
$\h$ the second peak at higher $k$ ($k_>$) becomes stronger and the first 
peak at smaller $k$ ($k_<$)
loses weight. At further higher $\h$ the small $k$ peak disappears.
It is to be noted that all the while both the peaks shift towards smaller
$k$-values, such that the mode structure can eventually repeat itself  at 
$\h\to \h+\pi$.  This behaviour is clearly visible in Fig.~\ref{phik}.
Note that the mode structure we obtain is due to boundary driving and thus  
 must vanish
in the thermodynamic limit of large $N$. We indeed find  $\tilde\phi_j^2\propto 1/N^2$ (Eq.~\eqref{eq:mode}). 

This mode behaviour in $\w-\h$ plane is shown in Fig.~\ref{phd}.
Here, the {\em line of crossover} is defined as
\bea
\tilde\phi_{j_1}^2(\omega,\h)=\tilde\phi_{j_2}^2(\omega,\h)
\eea
where $j_1,j_2$ denote the positions of the two maxima of $\tilde\phi_{j}^2(\omega,\h)$.
In general it is difficult to find this curve  analytically. 
However,  the limiting 
behaviour at large $\omega$ is easy to obtain. 
 Substituting $-\epsilon_j\equiv s$ ($0<s<4r$) we find
at large $\omega\, (\omega>>x)$ 
\bea
{N^2 \over 4 \a^2} \tilde\phi^2_s(\omega,\h)= \frac{s(4r-s)}{4r^2\omega^4}(\omega \cos \h-s \sin \h)^2\nn
\eea 
 which has two maxima at
\bea
s_{1,2}=\frac 14 \left[6r+\omega\cot \h\pm\sqrt{36r^2-4r\omega\cot \h+\omega^2 \cot ^2 \h}\right]. \nn
\eea
 Thus we find that  the above-mentioned crossover 
[$\tilde\phi^2_{s_1}(\omega,\h)=\tilde\phi^2_{s_2}(\omega,\h)$]
occurs at  (see Fig.~\ref{phd})
\bea
\tan \h= \f{\omega}{2r}. 
\eea

\section{ASEP with AC drive}

In this section we extend this study of periodic boundary drive to a system 
with asymmetric hop rate, {\it  i.e. }  $p \neq q$. 
Thus correlations are no longer negligible in time evolution of the density profile.
We assume that $p=r+\nu$ and $q=r-\nu$, with $\nu$ breaking the symmetry of 
the hopping rates.  
For small $\nu \ll r$ we can expand the local density and correlations 
in a perturbation series
\bea
\r_i
&=& \r + \sum_{n=0}^\infty \nu^n \r^{(n)}_i(t), \crcr
C_{i,j} &=& C + \sum_{n=0}^\infty \nu^n  C^{(n)}_{i,j}(t),
\eea
where $\r$ denotes the mean density and $C=\r (\r (N+1)-1)/N$ is the 
two-point correlation, for $\nu=0$ and in absence of AC driving (exact results for SEP). 
Note that $\trz_i(t)$ and $\tcz_{i,j}(t)$ denote time evolution in the SEP-system in
response to the AC drive, i.e., $\trz_i(t)=\phi_i(t)$  (as discussed in the previous section). 
The perturbation series lets us write down the time evolution at each order; 
up to the lowest order in $\nu$,
\bea
\frac{d\trz_i}{dt} &=& r \D_i\trz_i,\crcr
 \frac{d\trek_i}{dt} &=& r \D_i\trek_i +  \left(\trz_{i-1} - \trz_{i+1}\right)
-2 \left(\tcz_{i-1,i} - \tcz_{i,i+1}\right)\nn
\eea
and
\bea
\label{eq:corr}
  \frac{d\tcz_{i,j}}{dt} &=& r  \left( \D_i + \D_j\right) \tcz_{i,j}, \crcr
  \frac{d\tcz_{i,i+1}}{dt} &=& r \left( \tcz_{i-1,i+1} + \tcz_{i,i+2} -2\tcz_{i,i+1}\right),
\eea
where we have used the notation $\D_i g_{i,j}=g_{i+1,j}+g_{i-1,j}-2 g_{i,j}$.
The time evolution of correlations are non-local and thus in general difficult 
to solve. However, in this case, the following mean field solution turns out to exactly 
satisfy Eq.~\eqref{eq:corr} 
\beq
\tcz_{i,j} = \left. \r_i \r_j\right|_{\nu=0} = \r \left(\trz_i + \trz_j \right)+\trz_i \trz_j.
\label{mft-c}
\eeq
This simplifies the time evolution of first order perturbation in density to
\beq
  \frac{d\trek_i}{dt} = r \D_i\trek_i 
              + \left(1-2\r-2\trz_i\right)\left(\trz_{i-1} - \trz_{i+1}\right),
              \label{eq:ro1}
\eeq
where $\rho_i^{(0)}(t)=\phi_i(t).$ Since $\rho_i^{(1)}(t)$ must  satisfy the boundary conditions 
 $\rho_0^{(1)}(t)=0 = \rho_N^{(1)}(t)$, we use a   
discrete sine transformation :
\bea
\rho_i^{(1)}(t)= \sum_{j=1}^{N-1} \tilde \rho_j^{(1)}(t)  \psi_j(i). 
\eea
 where, as before, $\psi_j(i)= \sin (k_j i )$ with $j$ or $k_j=\pi j/N$ denoting the sine Fourier modes.
 Now, the set of equations in Eq. \eqref{eq:ro1} are decoupled, resulting  in
\bea
{d\tilde \rho_j^{(1)}(t) \over dt} = 
\begin{cases}
 \epsilon_j \tilde \rho_j^{(1)}(t) + (1- 2 \rho) \gamma_j  & j= odd \cr
\cr 
 \epsilon_j \tilde \rho_j^{(1)}(t)  - 2\mu_j & j= even
\end{cases}
\label{eq:rho1t}
\eea
where,
\bea
\gamma_j &=&  -{4 \over N} \sum_{l} \tilde \phi_l 
 {\sin {k_l} \sin {k_j} \over \cos {k_l} - \cos {k_j}} \\          
\mu_j &=& -{1 \over 2} \sum_{l} \tilde\phi_l \left[\tilde\phi_{j-l}
 + \tilde\phi_{j+l} \right]\sin {k_l} 
 \label{muj}.     
\eea
\begin{figure}[t]
 \includegraphics[width=7cm]{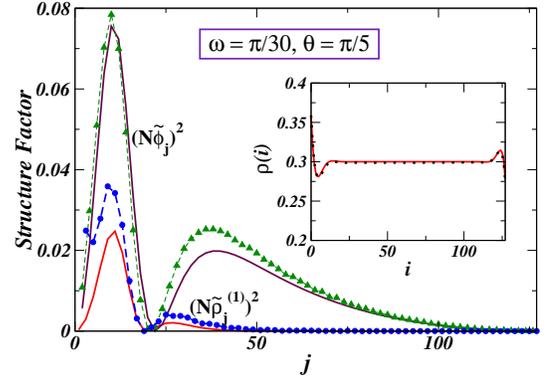}
\caption{(Color online) Density profile and structure factor of ASEP. 
The inset  compares  the  density profile  obtained from  Monte-Carlo simulations (symbols) with 
Eq. \eqref{eq:rho_i_asep} (solid line) for a system of size $N=128$  
with   $r=1/2$, asymmetry parameter of 
hopping rate $\nu=0.05$, and  mean density $\rho=0.3.$ The amplitude, frequency 
and phase of the AC boundary drives are $\alpha=0.1$,  $\o=\pi/30$ and $\theta=\pi/5$ respectively. 
The main figure shows normalized structure factor obtained from the sine transform of 
the  Monte-Carlo density profile (points); clearly  showing two branches. The odd (circle)  and 
the even (triangle) components are compared with  $(N\tilde \rho_j^{(1)})^2$ 
and  $(N\tilde \phi_j)^2$ respectively (Eq. \eqref{eq:rhojt1} and \eqref{eq:phit_j}). 
The curves for the odd component are scaled by a factor $7.5$ for visual clarity.}
  \label{fig:asep}
 \end{figure}
The sums  in  the above equations  extend only over even values of $l$ as $\tilde \phi_l=0$ for odd $l$ 
(see Eq. \eqref{eq:phit_j}). Note that  unlike $\gamma_j$,   which is proportional to $\alpha$,   $\mu_j$    varies  as $\alpha^2$ and 
can be neglected for $\alpha \ll 1$.  Thus,  in the small $\alpha$ limit   $\tilde \rho_j^{(1)}$  for  even $j$
vanishes exponentially as $t\to \infty$  (from Eq. \eqref{eq:rho1t}),  and for odd $j$  
 \bea
\tilde \rho_j^{(1)}(\omega,\theta)&=& {8 \alpha (1 -2 \rho)\over N^2( \epsilon_j^2+ \omega^2)} 
\sum_{l}  
{\sin^2{k_l} \over \epsilon_l^2+ \omega^2 } {\sin {k_j} \over \cos {k_l} 
- \cos {k_j}} \cr 
&& \left[ (\omega^2 - \epsilon_j \epsilon_l) \sin \theta 
- \omega (\epsilon_j +\epsilon_l) \cos \theta \right]. 
\label{eq:rhojt1}
\eea 
Finally, to the  linear order in $\nu$, the density profile is  given by 
\bea
\rho_i = \rho+
 \sum_{j=even}^{N-1} \tilde \phi_j \psi_j(i)  +
\nu \sum_{j=odd}^{N-1} \tilde \rho_j^{(1)} \psi_j(i).\label{eq:rho_i_asep}
\eea 
The  asymmetric bulk dynamics in ASEP  produces a drift 
in one preferable direction generating  the odd Fourier modes which were absent in SEP, 
whereas the even modes for small $\alpha$ is identical to that of SEP.  
In the inset of Fig. \ref{fig:asep} we compare the density profile 
calculated  from  Eq. \eqref{eq:rho_i_asep} with  that obtained from Monte-Carlo 
simulations for a system of size $N=128.$

Next, we investigate, whether  the structure factor of ASEP retains the 
bimodal feature   in  the $\o$ -$\theta$ plane.  It is sufficient to 
check that for only the odd Fourier modes, {\it i.e.,} $\tilde \rho_j^{(1)}$, 
as even modes remain unaltered for small $\alpha$. 
 Eq. \eqref{eq:rhojt1} shows that the dominant contributions in the
expression of $\tilde \rho_j^{(1)}$ come  from  
$l=j\pm1$ (remember that $j$  is an odd integer and $l$ takes only even values in the sum). 
Thus in the large $N$ limit $\tilde \rho_j^{(1)}$   vanishes at some  intermediate $k_j$, 
apart from  the two extreme values $k_j=0,\pi$ which satisfies 
$\tan \theta = 2 \o \e_j/(\o^2-\e_j^2).$  This indicates that  the  odd  component of the 
structure factor too has a bimodal structure. 

In Fig. \ref{fig:asep} we show  structure factors  
calculated  from  the  discrete sine  transform of the numerically obtained density profile 
(shown in  the inset). These structure factors, both for  odd and the even modes, agree 
qualitatively with  the  perturbation calculation 
Eq. \eqref{eq:phit_j} and \eqref{eq:rhojt1} (drawn with   lines). 
The small discrepancy between the theory and simulation may be attributed to 
keeping the perturbation
expansion up to first order, a mean field estimate of the correlation function ( Eq.~\eqref{mft-c}), 
and the linearization  leading to decoupling of modes (Eq.~\eqref{muj}).
In fact the  sinusoidal drive may force the system to dynamically 
cross the  boundaries between different phases 
(high density, low density  and maximal current phases) known to exist 
in ASEP driven by time-independent bias~\cite{pasep}.

In summary,  we observe  that  the density correlations  in ASEP do not destroy the 
bimodal structure. The  asymmetry in the  bulk drive for ASEP  generates {\it independent}  
odd modes in addition to the even modes present in the  structure factor of SEP.

 We close this section by commenting that
the   current across  $i$-th bond, up to first order in $\nu$, can be expressed as
\bea
J_{i,i+1} &=& r \left(\phi_i-\phi_{i+1} \right) + \nu  [ 2\r(1-\r) + (1-2 \r ) 
\left(\phi_i+\phi_{i+1} \right)    \cr 
&&  -2 \phi_i \phi_{i+1}  + r  ( \r_i^{(1)}  -  \r_{i+1}^{(1)})   ]. \nn
\eea
This is not equal on all bonds, both for SEP ($\nu=0$) and ASEP ($\nu \neq 0$), 
showing that the system always  remains far from steady state.
 
\section{Conclusion}
In this paper
we have shown emergence of an interesting bi-modality in the structure factor of
simple AC driven diffusive systems namely SEP and ASEP.  
Our main prediction is that in a diffusive system two modes of transport,
one a long wavelength mode and another a short wavelength mode, remains 
operative and their relative weight gets exchanged periodically in time.
This behaviour repeats modulo $\pi$
in $\h=\w t$.  The region of coexistence of the two dominating 
modes shrinks in the $\h-\w$ plane with increase in driving frequency $\w$.
 
This prediction can be directly tested in experiments on sterically stabilized
colloids confined in narrow glass channels. 
It was shown experimentally, that a system of hard core colloidal particles of diameter $\s$
confined within a narrow channel of width $< 2\s$ is describable by SEP~\cite{Wei2000}. 
Recent experiments have studied impact of directed external drive on colloids confined 
in narrow channel~\cite{Henseler2006}. In Ref.~\cite{Henseler2006} a DC drive on confined 
colloids were generated by tilting the table, so that  the gravitational force could be 
utilized to drag the colloids. Similar setup may be used to test our prediction, e.g., by 
simply oscillating the table by a small amplitude one may generate
an AC drive. Optical microscopy, or direct light scattering 
may be used to find structure factor of the system to test our predictions. 

\smallskip 

\acknowledgements The work of DC is part of the research program of the ``Stichting voor Fundamenteel
Onderzoek der Materie (FOM)'', which is financially supported by the ``Nederlandse
organisatie voor Wetenschappelijk Onderzoek (NWO)''. We thank MPI-PKS Dresden where 
this work was initiated.
\bibliography{ac}
\end{document}